\documentstyle[aps, prl, twocolumn, epsf, floats]{revtex}

\def\BEQ{\begin{eqnarray}}
\def\EEQ{\end{eqnarray}}
\def\BML{\begin{mathletters}}
\def\EML{\end{mathletters}}
\def\BE{\begin{equation}}
\def\EE{\end{equation}}

\draft 

\begin{document}
\title{Role of symmetry and dimension for pseudogap phenomena}
\author{S. Allen$^{1}$, H. Touchette$^{1}$, S. Moukouri$^{1}$, Y.M. Vilk$^{3}$, and
A.-M. S. Tremblay$^{1,2}$\cite{email}}
\address{$^{1}$D\'{e}partement de Physique and Centre de Recherche en Physique du\\
Solide and $^{2}$Institut canadien de recherches avanc\'{e}es}
\address{Universit\'{e} de Sherbrooke, Sherbrooke, Qu\'{e}bec, Canada J1K 2R1.\\
$^{3}$2100 Valencia Dr. apt. 406, Northbrook, IL 60062}
\date{April, revised August 1999}
\maketitle

\begin{abstract}
The attractive Hubbard model in $d=2$ is studied through Monte Carlo
simulations at intermediate coupling. There is a crossover temperature $%
T_{X} $ where a pseudogap appears with concomitant precursors of Bogoliubov
quasiparticles that are {\it not} local pairs. The pseudogap in $A\left( 
{\bf k}_{F},\omega \right) $ occurs in the renormalized classical regime
when the correlation length is larger than the direction-dependent thermal
de Broglie wave length, $\xi _{th}=\hbar v_{F}\left( {\bf k}\right) /k_{B}T.$
The ratio $T_{X}/T_{c}$ for the pseudogap may be made arbitrarily large when
the system is close to a point where the order parameter has $SO\left(
n\right) $ symmetry with $n>2$. This is relevant in the context of $SO\left(
5\right) $ theories of high $T_{c}$ but has more general applicability.
\end{abstract}


Normal state pseudogaps observed in angle-resolved photoemission experiments
(ARPES)\cite{ARPES} and tunneling\cite{Tunnel} have become the focus of much
of the research on high-temperature superconductors. Theoretically,
strong-coupling models of this phenomenon include repulsive doped insulator
models\cite{Lee}, and attractive models with preformed local pairs\cite
{Preformed,Randeria92,Stintzing}. At intermediate coupling, resonant pair
scattering has been invoked\cite{Levin}. In weak coupling, antiferromagnetic
or superconducting fluctuations also lead to pseudogap formation before long
range order is established\cite{VT,LongPaper,VA,Schmalian}. Finally,
arguments about the relevance of phase fluctuations\cite{PhaseFluct} do not
specify whether the small superfluid density comes from strong coupling or
from thermal fluctuation effects.

Although the effect of superconducting fluctuations on the density of states
was studied long ago\cite{1970}, the relevance of these critical
fluctuations on pseudogap phenomena in high temperature superconductors has
been questioned mainly because, even in low dimension, the critical region
should be quite small compared with the size of the pseudogap region
observed in high temperature superconductors. Indeed, even for the
Kosterlitz-Thouless (KT) transition in two dimensions, the range of
temperatures where superconducting phase fluctuations occur is of order $%
k_{B}T_{c}^{2}/E_{F}$ unless disorder depresses $T_{c}$\cite{Sharapov}.
Another question for the original critical-fluctuation calculations is that,
contrary to density of states pseudogaps, a pseudogap can appear in the
single-particle spectral weight $A\left( {\bf k}_{F}{\bf ,}\omega \right) $
only when the fluctuation-induced scattering rate at the Fermi surface $%
\mathop{\rm Im}%
\Sigma \left( {\bf k}_{F},\omega =0\right) $ increases with decreasing
temperature, a behavior that is opposite to that predicted by the
traditional phase-space arguments of Fermi liquid theory.

The last objection has already been answered\cite{VT} by showing that in $%
d=2 $ the Fermi liquid arguments fail when one enters the so-called
renormalized classical (RC) regime of fluctuations ($d=3$ is the upper
critical dimension \cite{VT}). This RC regime is the one where critical
slowing down leads to increasing dominance of classical fluctuations $\left(
\hbar \omega _{c}<k_{B}T\right) $ as temperature is lowered. However, the
question of the large temperature range where pseudogaps appear has to be
addressed theoretically both at the qualitative and at the quantitative
levels. On the quantitative side, the dependence of the critical region on
interaction strength and quasi two-dimensionality is not obvious, but it has
been argued \cite{Preosti} that specific models can give large critical
regions even at intermediate coupling. On the qualitative side, a factor
that may considerably enlarge the size of the pseudogap regime is the
proximity to a point where the order-parameter symmetry is $SO\left(
n\right) $ with $n>2$. Indeed, in $d=2$ the critical region can become much
larger when $n>2$ since then the Mermin-Wagner theorem implies that the
transition temperature is pushed down to $T=0$. We argue that both
conditions, namely $d=2$ and higher symmetry, are generic for
high-temperature superconducting materials. Indeed, in the underdoped
region, where the pseudogap is largest, these materials are highly
anisotropic (quasi two-dimensional) and it has been proposed that the order
parameter may have both antiferromagnetic and superconducting character
corresponding to approximate $SO\left( 5\right) $ symmetry\cite{Zhang}.

The attractive Hubbard model may be used to illustrate the properties of
pseudogaps that appear in such situations of approximate high-symmetry in $%
d=2$. For our purposes, the important characteristic of this model is that
the long-wavelength critical behavior is as follows for all values of
interaction $U$. At half-filling, there is a zero-temperature phase
transition that breaks the finite-temperature $SO\left( 3\right) $ symmetry 
\cite{NoteSO(4)} while away from half-filling, there is a KT transition at
finite temperature. The corresponding ground state breaks $SO\left( 2\right) 
$ symmetry$.$ While the details of this model are clearly inappropriate for
high-temperature superconductors, it is useful to illustrate a number of
general points that should be applicable to models with transition
temperatures that are pushed down from their mean-field value by a
combination of low dimension and high order-parameter symmetry $SO\left(
n>2\right) $\cite{Zhang}.

In this paper, we present Monte Carlo simulations for the $d=2$ Hubbard
model at $U=-4t,$ a value that is slightly on the BCS side of the BCS to
Bose-Einstein crossover curve $\left( U<U\left( T_{c}^{\max }\right) \right) 
$\cite{Singer98}. We use units where nearest-neighbor hopping is $t=1,$
lattice spacing is unity, $\hbar =1$ and $k_{B}=1$. Previous numerical work
charted the phase diagram\cite{Scalettar}. They have also investigated the
pseudogap phenomenon mostly in strong coupling where, we stress, the Physics
is different from the case discussed here\cite{Singer98,Singer96,Singer99}.
On the weak-coupling side of the BCS to Bose-Einstein crossover, there have
been numerical studies of KT superconductivity\cite{Moreo92} as well as
several discussions of pseudogap phenomena in the spin properties and in the
total density of states at the Fermi level\cite
{Randeria92,Trivedi95,Singer96}. The only study of $A\left( {\bf k},\omega
\right) $ was restricted to regions far from the $SO\left( 3\right) $
symmetric point\cite{Singer99}.

Here we establish a dynamical connection between the appearance of the RC
regime in the pairing collective modes and pseudogap formation in
single-particle quantities. In particular, we show that a) Close to a high
symmetry point, pseudogaps can appear at a crossover temperature $T_{X}$
that scales with the mean-field transition temperature while the real
transition may occur at much lower temperature, $T_{c},$ leading to a wide
temperature range for the pseudogap. b) At the crossover temperature, one
enters the RC regime where the characteristic frequency for fluctuations
becomes smaller than the temperature. c) Pseudogap in weak-to-intermediate
coupling do not require resonance in the two-particle correlations\cite
{NotePairPseud}. d) To have a pseudogap in $A\left( {\bf k}_{F}{\bf ,}\omega
\right) $ for a given wave vector, it is not enough to have the collective
mode (two-particle) correlation length satisfy $\xi $ $>1.$ It is necessary
that $\xi $ becomes larger than the single-particle thermal de Broglie
wavelength $\xi _{th}=v_{F}\left( {\bf k}\right) /T$. This implies in
particular that even for an isotropic interaction, as temperature decreases
a pseudogap opens first near the zone edge, where $v_{F}$ is small, and it
opens last along the zone diagonal where $v_{F}$ is largest. This anisotropy
would be amplified for an anisotropic interaction of $d-$wave type\cite
{Preosti}. For $U=-4$, the condition $\xi $ $>\xi _{th}$ is realized near
the zone edge for $\xi $ not so large. Analytical arguments for the above
results have appeared elsewhere\cite{LongPaper,VA,Vilk}.

Let us first recall a few results at half-filling, $\left\langle
n\right\rangle =1,$ where the chemical potential $\mu $ vanishes. There the
canonical transformation $c_{i\downarrow }\rightarrow \left( -1\right)
^{i_{x}+i_{y}}c_{i\downarrow }^{\dagger }$ maps the attractive model onto
the repulsive one at the same filling. The ${\bf q}=0,$ $s-$wave
superconducting fluctuations and the ${\bf Q}=\left( \pi ,\pi \right) $
charge fluctuations are mapped onto the three components of
antiferromagnetic spin fluctuations of the repulsive model and hence they
are degenerate. Because of this degeneracy, the order parameter at
half-filling has $SO\left( 3\right) $ symmetry\cite{NoteSO(4)}, hence, by
the Mermin-Wagner theorem, in two dimensions the phase transition is at $%
T_{c}=0$. Results for the attractive model are easily extracted from
simulations of the canonically equivalent repulsive model. The pair
structure factor $S_{\Delta }=\left\langle \Delta ^{\dagger }\Delta +\Delta
\Delta ^{\dagger }\right\rangle $ with $\Delta =\frac{1}{\sqrt{N}}%
\sum_{i=1}^{N}c_{i\uparrow }c_{i\downarrow }$ and the ${\bf Q=}\left( \pi
,\pi \right) $ charge structure factor $S_{c}=\left\langle \rho _{{\bf Q}%
}\rho _{-{\bf Q}}\right\rangle $ are identical, showing an increase as
temperature decreases and then size-dependent saturation. The sudden rise of 
$S_{\Delta }$ as temperature decreases indicates a crossover to a RC regime
with a concomitant opening of the pseudogap\cite{VT,LongPaper,Moukouri99} in 
$A\left( {\bf k}_{F,}\omega \right) $. The crossover temperature is a
sizeable fraction of the mean-field transition temperature.%
\begin{figure}%
%
\centerline{\epsfxsize 8.5cm \epsffile{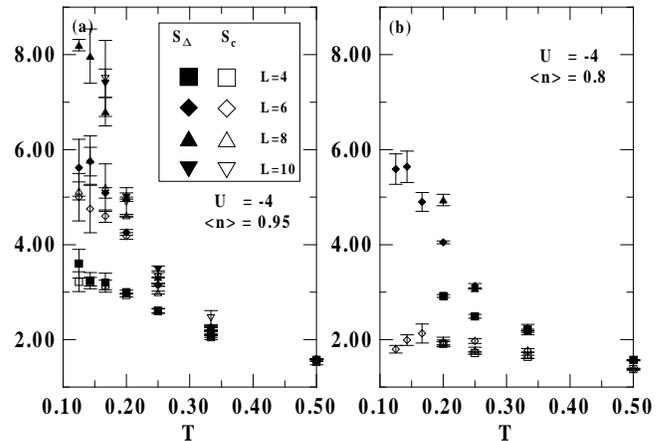}}%
%
\caption{Pair (filled symbols) and charge (open symbols) structure factors for $U=-4$
and various system sizes. (a) Filling
$\left\langle n \right\rangle=0.95$. (b) Filling $\left\langle n \right\rangle=0.8$.}%
%
\label{fig1}%
%
\end{figure}%
%
\quad

Slightly away from half-filling, the $SO\left( 3\right) $ symmetry is
formally broken by the chemical potential\cite{NoteSO(4)} since it couples
only to the charge part of the triplet. However, in the regime where the
temperature is larger than the symmetry-breaking field, the symmetry is
approximately satisfied.\cite{VA} For filling $\left\langle n\right\rangle
=0.95$, we have $T>\left| \mu \right| $ for the whole range of temperature
shown in Fig.1(a) ($\mu \sim -0.07$ at $T=0.125$)$.$ On this figure,
superconducting ($S_{\Delta },$ filled symbols) and charge ($S_{c},$ open
symbols) structure factors are of comparable size when one enters the RC
regime, showing that we have approximate $SO\left( 3\right) $ symmetry for
various sizes $L\times L.$ The beginning of the RC regime, that occurs at a
temperature nearly identical to the crossover temperature \cite{VT} $T_{X}$
identified at half-filling, is signaled by the increase in the magnitude of
correlations. Eventually, the concomitant increase of $\xi $ leads to the
size-dependence apparent at lower temperature in Fig.1(a). The plot in
Fig.1(a) resembles the result at half-filling.\cite{White89,LongPaper} The
near equality of superconducting and charge fluctuations at the crossover to
the RC regime should be contrasted with the case $\left\langle
n\right\rangle =0.8$ in Fig.1(b) where the charge fluctuations show
basically no critical behavior when the superconducting correlations begin
to do so. Hence, at this filling, there is little $SO\left( 3\right) $
symmetry left at $T_{X}.$ This is expected since the symmetry-breaking field 
$\left| \mu \left( T_{X}\right) \right| =\left| \mu \left( 0.25\right)
\right| =0.26$ is comparable to $T_{X}$. One basically enters directly into
the RC regime of a $SO\left( 2\right) $ KT transition\cite{Scalettar}. In
this regime $dT_{c}\left( n\right) /dn>0.$

Let us go back to the filling $\left\langle n\right\rangle =0.95.$ At this
filling, it has been estimated\cite{Scalettar} that $T_{c}<0.1$ and $%
dT_{c}\left( n\right) /dn<0$. We already showed that at this filling, one
has approximate $SO\left( 3\right) $ symmetry. One can confirm that the
increase of $S_{\Delta }$ and $S_{c}$ at $T_{X}$ comes from RC fluctuations
by calculating the spectral weight for superconducting fluctuations $\chi
_{\Delta }^{\prime \prime }$ (imaginary part of the $T-matrix$), $\chi
_{\Delta }^{\prime \prime }\left( \omega \right) =\int dt\frac{1}{2}%
\left\langle \left[ \Delta \left( t\right) ,\Delta ^{\dagger }\right]
\right\rangle e^{i\omega t}.$ The even part\cite{NotePaire} of $\chi
_{\Delta }^{\prime \prime }\left( \omega \right) /\omega $ was obtained from
a Monte Carlo calculation of the imaginary-time quantity $\left\langle
\Delta \left( \tau \right) \Delta ^{\dagger }+\Delta ^{\dagger }\left( \tau
\right) \Delta \right\rangle $ followed by Maximum Entropy inversion\cite
{Meshkov}. The even part of $\chi _{\Delta }^{\prime \prime }\left( \omega
\right) /\omega $ is plotted on Fig.2 for an $8\times 8$ system and various
temperatures.%
\begin{figure}%
%
\centerline{\epsfxsize 4cm \epsffile{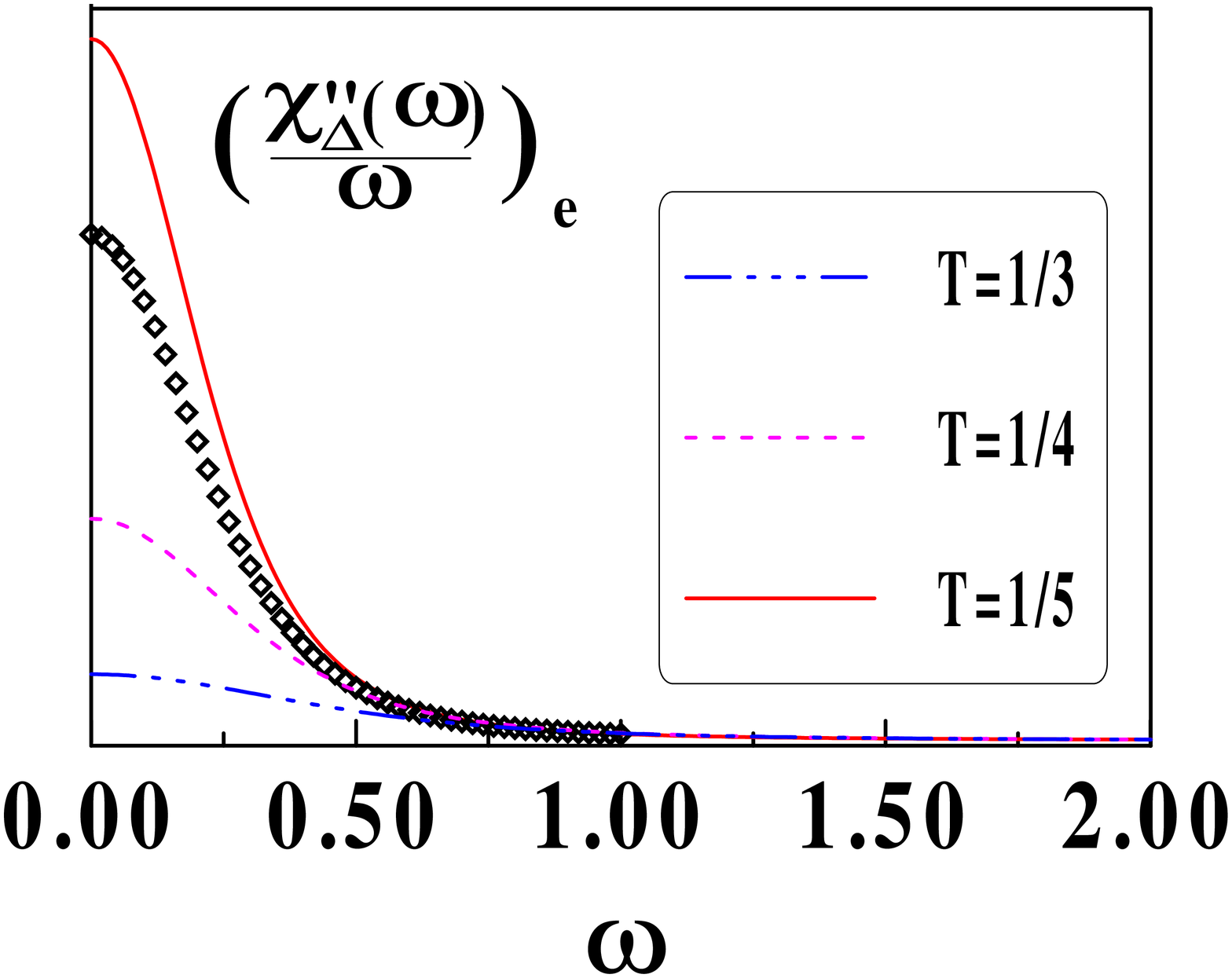}}%
%
\caption{Even part of the pair spectral weight $\chi''_\Delta(\omega)/\omega$ 
obtained from analytic continuation
of imaginary time Monte Carlo data for $U=-4$, $\Delta\tau=1/10$, 
$\left\langle n \right\rangle=0.95$ and size $8 \times 8$, except for symbols that are 
for $6\times 6$ at $T=1/5$. About $10^5$ sweeps 
were done in each case.}%
%
\label{fig2}%
%
\end{figure}%
%
The symbols are for a $6\times 6$ system at $T=1/5.$ The maximum is at zero
frequency, as for an overdamped mode $\chi _{\Delta }^{-1}\left( {\bf q}%
,\omega \right) \propto \xi ^{-2}$ $+q^{2}-i\omega /\omega _{0}$. The
characteristic frequency, given by the half-width at half maximum, is $%
\omega _{c}=\omega _{0}/\xi ^{2}$ with $\omega _{0}$ a microscopic
relaxation rate. There is a marked narrowing of the width in frequency as
temperature decreases. One enters the RC regime when $\omega _{c}\simeq
1/4\simeq T_{X}$, a temperature larger than $T_{c}\left( <0.1\right) $. At $%
T=1/5$, the correlation length is becoming comparable with system size since
the $6\times 6$ system gives a result that differs from $8\times 8$.

The effect of RC collective fluctuations on single-particle quantities is
illustrated in Fig.3 (a) to (c) that show density plots of the
single-particle spectral weight $A\left( {\bf k,}\omega \right) $ for an $%
8\times 8$ system at, respectively, $T=1/3,1/4,$ and $1/5.$%
\begin{figure}%
%
\centerline{\epsfxsize 6cm \epsffile{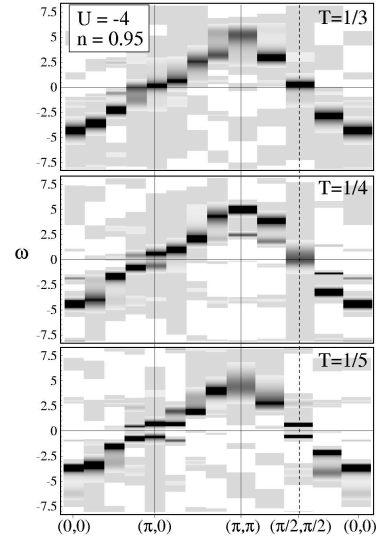}}%
%
\caption{Density plot of the single-particle spectral weight 
$A({\bf k},\omega)$ from about $10^5$ Monte Carlo 
sweeps for $U=-4$, $\Delta\tau=1/10$, 
$\left\langle n \right\rangle=0.95$ and size $8 \times 8$. 
Top to bottom, $T=1/3, 1/4, 1/5$. 
The dilation by $\protect\sqrt(2)$ of the axis from $(\pi,\pi)$ to $(0,0)$ 
allows a comparison of Fermi velocities.}%
%
\label{fig3}%
%
\end{figure}%
%
$\quad $When temperature reaches $T=1/4$ one notices that a minimum ({\it %
pseudogap}) around ${\bf k}=\left( 0,\pi \right) $, $\omega =0$ develops
along with two maxima away from $\omega =0$. The latter maxima are {\it %
precursors} of the Bogoliubov quasiparticles of the ordered state\cite
{LongPaper}\cite{Kyung}. The pseudogap becomes deeper and deeper as
temperature decreases, the distance between maxima remaining about constant,
as observed in high temperature superconductors\cite{Tunnel}. The condition
for the appearance of a pseudogap in $A\left( {\bf k,}\omega \right) $ is
not only that we should be in the RC regime and in low dimension but also
that $\xi $ $/\xi _{th}$ should be large\cite{VT,LongPaper}. This
illustrated by the fact that the pseudogap at ${\bf k=}\left( \pi /2,\pi
/2\right) ,$ where the Fermi velocity is larger, is not opened yet at $T=1/4$
despite the fact that at $T=1/5$ the pseudogap is of comparable size both
around ${\bf k}=\left( 0,\pi \right) $ and at ${\bf k=}\left( \pi /2,\pi
/2\right) $, in concordance with the equality of the gaps at these two
points in the zero-temperature spin-density wave state. From the slopes in
Figs.3 (a) to (c), $v_{F}$ is clearly larger at $\left( \pi /2,\pi /2\right) 
$ meaning that the condition $\xi $ $>\xi _{th}=v_{F}/T$ is harder to
satisfy at this wave vector. Numerical estimates show that $\xi $ is nearly
isotropic by contrast with $v_{F}.$ These estimates are consistent with the
appearance of the pseudogap in $A\left( {\bf k,}\omega \right) $ when $\xi $ 
$\sim \xi _{th}.$

As in any numerical simulation, finite-size effects should be considered
carefully. There are two important intrinsic lengths in this problem, namely 
$\xi _{th}$ and $\xi .$ When $L\ll \xi _{th},$ the system acts as if it was
basically in the quantum zero-temperature limit of a finite system. We have
checked that at $T=1/8,$ $A\left( {\bf k},\omega \right) $ shows real gaps,
instead of pseudogaps, that appear at progressively higher temperature in
systems of smaller size. For $T=1/3,1/4$ on the other hand, estimates of $%
\xi _{th}$ and of $\xi $ as well as calculations for $6\times 6$ systems and
a few $10\times 10$ systems suggest that our numerical results for $A\left( 
{\bf k,}\omega \right) $ on $8\times 8$ systems are free of appreciable size
effects, i.e. $L$ $>\xi ,\xi _{th}$. This can also be checked by the size
dependence of the results in Fig.1(a) and Fig.2. While size effects become
important in $\chi _{\Delta }^{\prime \prime }\left( \omega \right) /\omega $
when $\xi $ exceeds $L$, as long as $\xi _{th}<L$ then it is possible to see
thermally induced pseudogap effects in $A\left( {\bf k},\omega \right) $
even if $\xi >L$\cite{VT}.

In summary, the qualitative phase diagram for the attractive Hubbard model
in $d=2$ sketched in Fig.4 shows that near a point with high order parameter
symmetry, the transition temperature decreases while the pseudogap
temperature increases along with the mean-field transition temperature and
the zero-temperature gap.%
\begin{figure}%
%
\centerline{\epsfxsize 5.5cm \epsffile{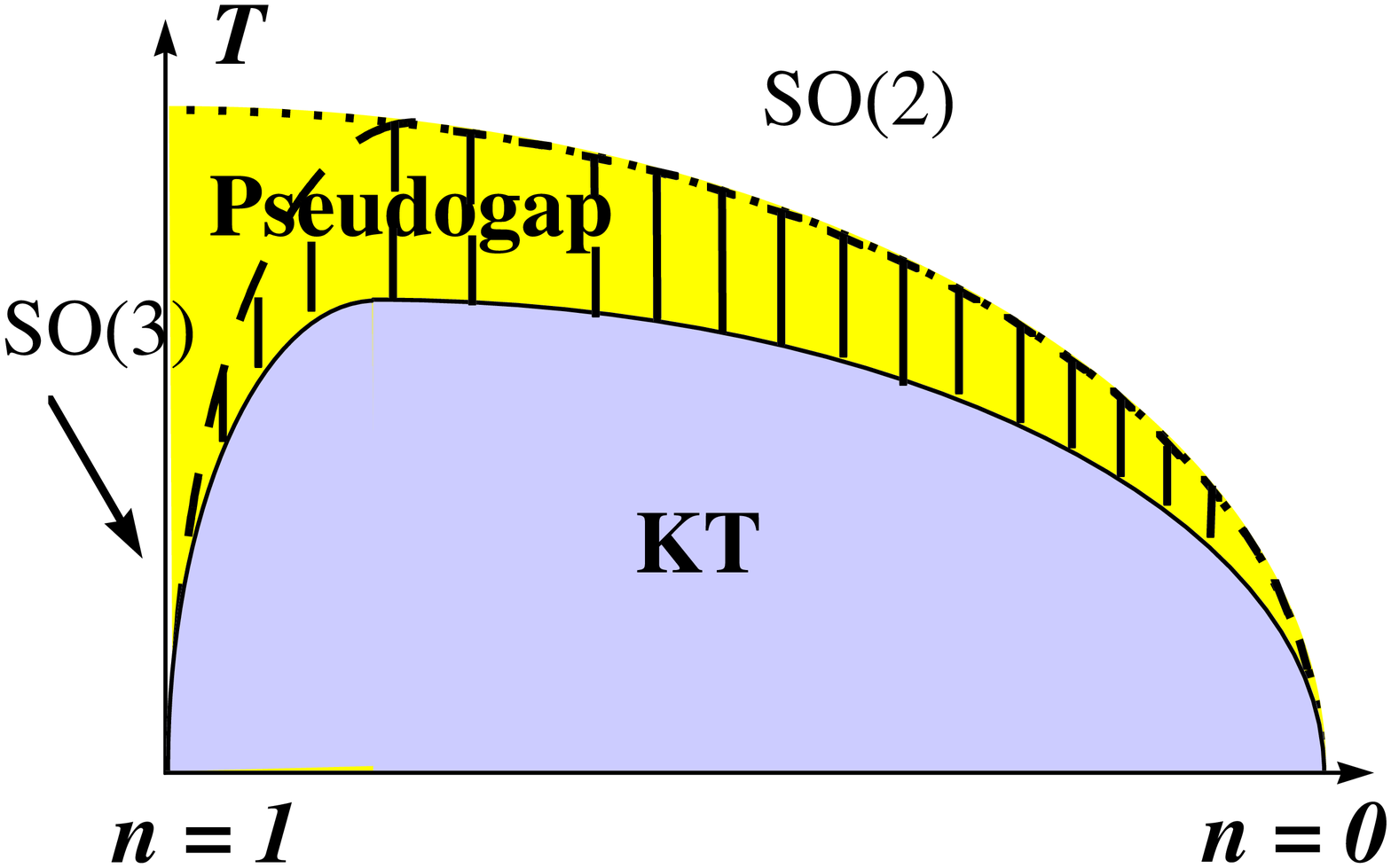}}%
%
\caption{Schematic crossover diagram for the $d=2$ attractive 
Hubbard model in the weak-coupling regime. The shaded area above
the KT phase, up to the dotted, line is the RC pseudogap regime.
The KT critical regime is hatched.}%
%
\label{fig4}%
%
\end{figure}%
%
\quad In the region where $dT_{c}/dn<0,$ the crossover to KT critical
behavior occurs in the RC pseudogap regime when $T$ is less than the
symmetry breaking field. In our simulations we would need larger system size
to reach the KT critical regime. Contrary to the scenario of Ref.\cite{Levin}%
, in our case dimension and symmetry contribute to create a wide pseudogap
region, there is no critical coupling strength, and furthermore one enters
the RC regime without sharp resonance in $\chi _{\Delta }^{\prime \prime
}\left( \omega \right) /\omega .$ Also, the precursors of Bogoliubov
quasiparticles in $A\left( {\bf k},\omega \right) $ occur under conditions
very different from those for strong-coupling Cooper pairs that are local
and do not need low dimension or large $\xi /\xi _{th}$. Comparisons with
non-perturbative many-body calculations should appear elsewhere\cite{Kyung}.

In high $T_{c}$ superconductors the competition is between
antiferromagnetism and superconductivity. Recent time-domain transmission
spectroscopy experiments\cite{Orenstein98} suggest that the RC regime for
the KT transition (hatched region in Fig.4) has been observed. Close enough
to the transition there is dimensional crossover to $d=3$. For
antiferromagnetic fluctuations, there are suggestions from NMR of a RC
regime \cite{Pines}, but there is no definite proof.

We acknowledge stimulating discussions with B. Kyung, R.J. Gooding, R.B.
Laughlin, J. Orenstein, D. S\'{e}n\'{e}chal, S.C. Zhang and participants of
the 1998 Aspen Summer Workshop on High-Temperature Superconductivity. We
thank H.-G. Mattutis for discussions on Monte Carlo methods, and D. Poulin
and L. Chen for numerous contributions to codes. Monte Carlo simulations
were performed in part on an IBM-SP2 at the Centre d'Applications du Calcul
Parall\`{e}le de l'Universit\'{e} de Sherbrooke. This work was partially
supported by NSERC (Canada), and by FCAR (Qu\'{e}bec).

\end{document}